\documentclass[prd,superscriptaddress,amsfonts,amssymb,onecolumn,amsmath,showpacs]{revtex4-2}
\usepackage{bm}
\usepackage{amsfonts}
\usepackage{tensor}

\usepackage{latexsym}
\usepackage[utf8]{inputenc}
\usepackage{graphicx}
\usepackage{amsmath}
\usepackage{palatino}
\usepackage{mathpazo}
\usepackage[british]{babel}
\usepackage{hhline}
\usepackage{multirow}
\usepackage{textcomp}
\usepackage{float}
\usepackage{booktabs}
\usepackage{dcolumn}
\usepackage{lipsum} 
\usepackage{mdframed}
\usepackage[]{mdframed}
\usepackage{hhline}
\usepackage{multirow}
\usepackage{ragged2e}
\usepackage{hyperref}
\hypersetup{colorlinks,citecolor=blue}
\hypersetup{colorlinks=true,linkcolor=red,filecolor=magenta,    urlcolor=cyan}
\usepackage{amsmath}
\usepackage{xcolor}
\usepackage{orcidlink}
\usepackage{epsfig}
\usepackage{caption}
\usepackage{subcaption}
\usepackage{commath}

\def\jnl@style{\it}
\def\aaref@jnl#1{{\jnl@style#1}}

\def\aaref@jnl#1{{\jnl@style#1}}

\def\aj{\aaref@jnl{AJ}}                   
\def\apj{\aaref@jnl{ApJ}}                 
\def\apjl{\aaref@jnl{ApJ}}                
\def\apjs{\aaref@jnl{ApJS}}               
\def\apss{\aaref@jnl{Ap\&SS}}             
\def\aap{\aaref@jnl{A\&A}}                
\def\aapr{\aaref@jnl{A\&A~Rev.}}          
\def\aaps{\aaref@jnl{A\&AS}}              
\def\mnras{\aaref@jnl{Mon.~Not.~Roy.~Astron.~Soc.}}             
\def\prd{\aaref@jnl{Phys.~Rev.~D}}        
\def\prc{\aaref@jnl{Phys.~Rev.~C}}  
\def\prl{\aaref@jnl{Phys.~Rev.~Lett.}}    
\def\qjras{\aaref@jnl{QJRAS}}             
\def\skytel{\aaref@jnl{S\&T}}             
\def\ssr{\aaref@jnl{Space~Sci.~Rev.}}     
\def\zap{\aaref@jnl{ZAp}}                 
\def\nat{\aaref@jnl{Nature}}              
\def\aplett{\aaref@jnl{Astrophys.~Lett.}} 
\def\apspr{\aaref@jnl{Astrophys.~Space~Phys.~Res.}} 
\def\physrep{\aaref@jnl{Phys.~Rep.}}      
\def\physscr{\aaref@jnl{Phys.~Scr}}       
\def\commat{\aaref@jnl{Comm.~Math.~Phys.}}              
\def\science{\aaref@jnl{Science}}               
\def\cqg{\aaref@jnl{Classical Quant.~Grav.}}            
\def\jpcs{\aaref@jnl{JPCS}}                                     
\def\ijmpd{\aaref@jnl{Int.~J.~Mod.~Phys.~D}}                    
\def\grg{\aaref@jnl{Gen.~Relat.~Gravit.}}               
\def\rpp{\aaref@jnl{Rep.~Prog.~Phys.}}          
\def\npa{\aaref@jnl{Nucl.~Phys.~A}}        
\def\lrr{\aaref@jnl{Living Rev.~Rel.}}                   
\def\jcap{\aaref@jnl{J.~Cosmology Astropart.~Phys.}}    
\def\rmp{\aaref@jnl{Rev.~Mod.~Phys.}}   
\def\epjc{\aaref@jnl{Eur.~Phys.~J.~C}} 
\def\plb{\aaref@jnl{~Phy.~Lett.~B}} 
\def\mpla{\aaref@jnl{Mod.~Phy.~Lett.~A}} 
\def\arxiv{\aaref@jnl{arxiv.org}}

\allowdisplaybreaks[1]
\renewcommand{\arraystretch}{1.1}
\begin{document}

\color{black}       

\title{Dynamical analysis of the covariant $f(Q)$ gravity models}

\author{S. A. Narawade\orcidlink{0000-0002-8739-7412}}
\email{shubhamn2616@gmail.com}
\affiliation{Department of Mathematics \& Statistics, Vishwakarma University, Pune-411048, Maharashtra, India}

\author{S. A. Kadam\orcidlink{0000-0002-2799-7870}}
\email{siddheshwar.kadam@dypiu.ac.in;
\\k.siddheshwar47@gmail.com}
\affiliation{Centre for Interdisciplinary Studies and Research, D Y Patil International University, Akurdi, Pune-411044, Maharashtra, India}

\begin{abstract}

\textbf{Abstract:} In this study, we explore the cosmological evolution of the Universe in the framework of covariant $f(Q)$ gravity, with a coupling function that evolves dynamically in proportion to the Hubble parameter. Two specific forms of the function are examined: a power-law model and a logarithmic model. By rewriting the cosmological field equations as an autonomous dynamical system, we determine and classify the corresponding critical points and analyze their stability. Our results show that both models are able to reproduce the sequence of cosmic evolution, including radiation, matter, and dark energy-dominated eras, along with the transitions between them. The physical properties at each critical point are described using key cosmological quantities such as the total EoS parameter, density parameters, and the deceleration parameter. The stability of the non-hyperbolic critical point is analyzed through center manifold theory. In addition, we present phase space trajectories along with the stability behavior of each critical point. The evolution plots for the density parameters of radiation, matter, and dark energy, along with the EoS parameter for the total, are illustrated for further analysis. Overall, the analysis suggests that the $f(Q)$ models considered here, within the context of covariant formulation, provide a consistent description of cosmic evolution and offer a promising approach to explaining the late-time acceleration of the Universe.  
\end{abstract}

\maketitle


\section{Introduction}\label{Introduction}

Recent observations in cosmology, including findings from the Dark Energy Spectroscopic Instrument (DESI), type Ia supernovae (SNIa) ~\cite{Riess:1998cb, Perlmutter:1998np}, large-scale structure (LSS)~\cite{Koivisto:2005mm, Daniel:2008et,Sokoliuk2023}, the Wilkinson Microwave Anisotropy Probe (WMAP)~\cite{WMAP:2003zzr,WMAP:2012nax}, the cosmic microwave background (CMB) and baryonic acoustic oscillations (BAO)~\cite{SDSS:2005xqv, SDSS:2009ocz}, imply that the conventional Cold Dark Matter (CDM) model may not be entirely sufficient. Several cosmological discrepancies, like the Hubble tension~\cite{Scolnic:2024hbh} and irregularities in structure formation~\cite{Nunes:2018xbm}, suggest that General Relativity (GR) may need adjustments to comprehensively account for cosmic development. One promising avenue of exploration is the geometric trinity of gravity~\cite{BeltranJimenez:2019esp}, which presents GR through three equivalent representations: curvature ($R$), torsion ($T$) and nonmetricity ($Q$).  However, these modifications differ at subsequent levels. Amongst them, $f(Q)$ gravity~\cite{Jimenez:2017tkx} has recently emerged as a particularly intriguing option. This theory alters GR while maintaining second-order field equations. It operates in the realm of symmetric teleparallel gravity, where both curvature and torsion are absent and gravity arises solely from nonmetricity. Notably, the field equations in both $f(T)$, $f(Q)$ theories are second-order instead of fourth-order~\cite{Ferraro:2006jd,Cai:2015emx,Bengochea:2008gz,Duchaniya:2022rqu,Jimenez:2017tkx,Bahamonde:IMPREV}, giving them a distinct advantage over $f(R)$~\cite{Starobinsky:2007,Sotiriou:2008rp} gravity theory. As a result, the evolution of $f(Q)$ theory provides a unique foundation for investigating various modified gravity models. The $f(Q)$ approach generalizes symmetric teleparallel gravity by substituting the geometric variable $Q$ in the Lagrangian of symmetric teleparallel gravity with a more general function $f(Q)$. It employs a connection that is both curvatureless and torsionless to establish the covariant derivative, in contrast to the Levi-Civita connection utilized in GR~\cite{Jarv:2018bgs}. The basis of this $f(Q)$ theory is nonmetricity $Q$, which provides a geometric explanation for how the length of a vector can change during parallel transport. Recent progress in $f(Q)$ gravity has led to exciting applications and captivating cosmic phenomena at the background level.

 The autonomous systems in the $f(Q)$ formalism for exponential, logarithmic and power-law models have been investigated in \cite{Vishwakarma2023,Bohmer2023} and analysis at infinity for the power-law model is presented in \cite{Paliathanasis2023}. The anisotropic locally rotationally symmetric (LRS) Bianchi type-I (LRS-BI) spacetime has been investigated in $f(Q)$ environment, contributing to the understanding of its unique properties and implications in cosmological models \cite{Sarmah2024}. The signatures in $f(Q)$ gravity are investigated for the prediction of the luminosity distance for gravitational waves compared to the standard electromagnetic one \cite{Frusciante2021}. The variation of the effective gravitational coupling has been analyzed in $f(Q)$ gravity, showing how this can be constrained using astrophysical observations \cite{dutta2025}. The non-flat cosmology in the $f(Q)$ formalism has been investigated in \cite{Shabani2024}. The phantom dark energy models in light of different cosmological observational data sets, such as Hubble and Pantheon+SHOES, using MCMC analysis have been investigated in \cite{Narawade2023}. The investigation of black holes, along with the construction of perturbative corrections to the Schwarzschild solution, focusing on different formulations, $f(Q)$ is investigated \cite{DAmbrosio2022}. The conservation law to test the viability of the $f(Q)$ models is tested in \cite{De2023}. The investigation of cosmic structures has yielded significant findings on the root-mean-square of matter fluctuations ($f_{\sigma 8}$), offering valuable insights \cite{Albuquerque:2022eac}. The success of the $f(Q)$ gravity theory in investigating black holes, wormholes, and the observational cosmic tensions is reviewed in \cite{Heisenberg2024}. This formalism is further extended, involving the trace of the energy-momentum tensor $f(Q, T)$~\cite{Xu:2019sbp}, the boundary term $f(Q, C)$~\cite{De:2024fQC}, and the matter Lagrangian $f(Q, L_m)$~\cite{Hazarika:2024alm}.

One of the significant challenges in theories of gravity stems from the nonlinearity of the field equations, which complicate the search for solutions, both analytical and numerical, and make it difficult to align predictions with observational data \cite{Narawade:2022a}. To tackle these complexities, researchers utilize a technique called "dynamical system analysis," which helps to solve these equations and manage the overall dynamic behavior of various systems. This approach seeks to uncover numerical solutions that provide deeper insights into the qualitative behavior of particular physical systems \cite{Bohmer2023}. A central idea in dynamical system analysis involves identifying critical points within a set of first-order ordinary differential equations. By computing the Jacobian matrix at these points and analyzing its eigenvalues, we can establish stability criteria \cite{Paliathanasis2023}. Once the critical points and their corresponding eigenvalues are determined, the system can be linearized around each point to study the flow behaviour in that region. This analysis emphasizes understanding the stability characteristics near specific critical points \cite{Perko2000}. In our study, we apply this framework of dynamical system analysis to examine different cosmic epochs. We set up the system for two well-motivated forms: one modeled by a power law and the other characterized by a logarithmic representation of $f(Q)$. These models are investigated at infinity and for different connections \cite{Vishwakarma2023, Paliathanasis2023, dutta2025}. We explored the dynamics of these models in the particular choice of flat connections. We investigated the cosmological parameters, like standard density parameters, for different cosmological epochs, like radiation, matter, and dark energy (DE), which are investigated at each critical point. The dynamical parameters, like the EoS and the behavior of the deceleration parameter, have been investigated. The dynamical system analysis has previously been investigated in $f(Q)$ and $f(Q, T)$ gravity within the coincident gauge framework. Notably, the vacuum scenario, where matter contributions are neglected, has been studied by \cite{Paliathanasis2023} without imposing the coincident gauge. In this work, we extend the existing vacuum dynamical analysis to a non-vacuum framework by explicitly incorporating matter and radiation components within coincident $f(Q)$ gravity. This generalization enables a more comprehensive phase-space description. Furthermore, the inclusion of matter fields offers deeper insight into the interplay between geometry and cosmic fluids in covariant $f(Q)$ gravity.

This work is organized as Section \ref{formalism_Section} presents detailed information on covariant $f(Q)$ gravity field equations. Section \ref{DSA_Section} presents detailed formation and analysis of autonomous dynamical systems for power-law model and logarithmic model. Also the stability of the non-hyperbolic critical point has been analyzed using center manifold theory. Section \ref{conclusion} concludes the important findings and remarks of the present work.

\section{Mathematical Formalism}\label{formalism_Section}
We begin with the action of $f(Q)$ gravity \cite{Jimenez:2017tkx}, 
\begin{equation}\label{AE}
    S = \int \frac{1}{2\kappa}f(Q)\sqrt{-g}~d^{4}x + \int \mathcal{L}_{m}\sqrt{-g}~d^{4}x~,
\end{equation}
where $f(Q)$ is an arbitrary function of the nonmetricity scalar $Q$ and $\mathcal{L}_{m}$ is the matter Lagrangian. The nonmetricity scalar is defined as $Q = Q_{\lambda\mu\nu}P^{\lambda\mu\nu}$, where the nonmetricity tensor $Q_{\lambda\mu\nu}$ equipped with the metric tensor $g_{\mu\nu}$ and the affine connection $\Gamma^{\lambda}_{~~\mu\nu}$ is given as \cite{Beh:2021wva, Zhao:2021zab, Narawade:2025},
\begin{eqnarray}\label{NMS}
    Q_{\lambda\mu\nu} &=& \nabla_{\lambda}g_{\mu\nu} = \partial_{\lambda}g_{\mu\nu}-\Gamma^{\alpha}_{~~\lambda\mu}g_{\alpha\nu}-\Gamma^{\alpha}_{~~\lambda\nu}g_{\alpha\mu}~,
\end{eqnarray}
and its conjugate, called the super-potential, is given by,
\begin{equation}\label{SP}
   P^{\lambda}_{~~\mu\nu} = -\frac{1}{4}Q^{\lambda}_{~\mu \nu} + \frac{1}{4}\left(Q^{~\lambda}_{\mu~\nu} + Q^{~\lambda}_{\nu~~\mu}\right) + \frac{1}{4}Q^{\lambda}g_{\mu \nu}- \frac{1}{8}\left(2 \tilde{Q}^{\lambda}g_{\mu \nu} + {\delta^{\lambda}_{\mu}Q_{\nu} + \delta^{\lambda}_{\nu}Q_{\mu}} \right)~,
\end{equation}
with the traces of nonmetricity as $Q_{\alpha} = Q_{\alpha~~\mu}^{~~\mu}$ and $\tilde{Q}_{\alpha}=Q^{\mu}_{~\alpha\mu}$. An alternative definition, $Q = -Q_{\lambda\mu\nu} P^{\lambda\mu\nu}$, has been introduced in the literature, which effectively flips the sign of the nonmetricity scalar $Q$. This difference in convention is important when comparing results across various $f(Q)$ studies. In the framework of symmetric teleparallel gravity, replacing the Ricci scalar $R$ in the Einstein-Hilbert action with the nonmetricity scalar $Q$ leads to the symmetric teleparallel equivalent of GR (STEGR). However, like standard GR, STEGR does not resolve the dark energy and dark matter problems. To address these issues, modified gravity models based on a generalized function $f(Q)$ have been proposed, analogous to the well-known $f(R)$ extensions of GR.\\

\noindent By varying the action Eqn. \eqref{AE} with respect to the metric tensor, we can obtain the field equation
\begin{equation}\label{FE}
    \frac{2}{\sqrt{-g}}\nabla_{\lambda}\left(\sqrt{-g}f_{Q}P^{\lambda}_{~~\mu\nu}\right) - \frac{1}{2}g_{\mu \nu}f + f_{Q}(P_{\mu\lambda\alpha}Q^{~~\lambda \alpha}_{\nu} - 2Q_{\lambda \alpha \mu}P^{\lambda \alpha}_{~~~\nu}) = \kappa T_{\mu \nu}~.
\end{equation}
Utilizing the covariant formulation, which has been developed and effectively applied to the study of geodesic deviation and cosmological phenomena, we can rewrite the above field equation as
\begin{equation}\label{NFE}
    f_{Q}\overcirc{G}_{\mu\nu}+\frac{1}{2}g_{\mu\nu}(Qf_{Q}-f)+2f_{QQ}P^{\lambda}_{~~\mu\nu}\overcirc{\nabla}_{\lambda}Q = \kappa T_{\mu \nu}~,
\end{equation}
Here, $\overcirc{\nabla}$ denotes the covariant derivative defined with respect to the Levi-Civita connection. The quantity $f_Q$ represents the derivative of the function $f$ with respect to the nonmetricity scalar $Q$. The tensor $\overcirc{G}{\mu\nu}$ is given by $\overcirc{G}{\mu\nu} = R_{\mu\nu} - \frac{1}{2} g_{\mu\nu} R$, where $R_{\mu\nu}$ and $R$ correspond to the Ricci tensor and Ricci scalar, respectively, both constructed from the Levi-Civita connection. In the special case where $f(Q)$ is linear in $Q$, the field equations reduce to those of GR. Furthermore, by performing a variation of Eq. \eqref{AE} with respect to the affine connection, one obtains the corresponding equation of motion associated with the nonmetricity scalar.
\begin{equation}\label{eq7}
    \nabla_{\mu}\nabla_{\nu}\left(\sqrt{-g}f_{Q}P^{\mu\nu}_{~~~\lambda}\right)=0.
\end{equation}
\noindent The affine connection considered here represents only a special case within one of the three classes introduced in the following discussion. Therefore, it is necessary to move beyond this particular gauge choice. To analyze the background geometry, we adopt a spatially homogeneous, isotropic and flat FLRW spacetime.
\begin{equation}\label{eq8}
    ds^2=-dt^2+a(t)^2\left[dr^{2}+r^{2}\left(d\theta^{2}+sin^{2}\theta ~d\phi^{2}\right)\right],
\end{equation}
where $a(t)$ is the scale factor and the Hubble function defined as $H= \frac{\dot{a}}{a}$ with $\dot{a}$ is derivative of $a$ with respect to time $t$. The connections component can be systematically derived as,
\begin{eqnarray}\label{CNS}
\Gamma^{t}{}_{\nu\rho} &=&
{\renewcommand{\arraystretch}{1.5}
\begin{pmatrix}
~~\mathcal{K}_{1}~~ & ~~0~~ & ~~0~~ & 0 \\
~~0~~ & ~~\mathcal{K}_{2}~~ & ~~0~~ & 0 \\
~~0~~ & ~~0~~ & ~~\mathcal{K}_{2} r^{2}~~ & 0 \\
0 & 0 & 0 & \mathcal{K}_{2} r^{2}\sin^{2}\theta
\end{pmatrix}}~, \quad\quad\quad  
\Gamma^{r}{}_{\nu\rho} =
{\renewcommand{\arraystretch}{1.5}
\begin{pmatrix}
~~0~~ & ~~\mathcal{K}_{3}~~ & 0 & 0 \\
~~\mathcal{K}_{3}~~ & 0 & 0 & 0 \\
0 & 0 & ~~-r~~ & 0 \\
0 & 0 & 0 & -r\sin^{2}\theta
\end{pmatrix}}~,\nonumber\\[10pt]
\Gamma^{\theta}{}_{\nu\rho} &=&
{\renewcommand{\arraystretch}{1.5}
\begin{pmatrix}
0 & 0 & ~~\mathcal{K}_{3}~~ & 0 \\
0 & 0 & ~~\frac{1}{r}~~ & 0 \\
~~\mathcal{K}_{3}~~ & ~~\frac{1}{r}~~ & 0 & 0 \\
0 & 0 & 0 & -\sin\theta\cos\theta
\end{pmatrix}}~, \quad\quad\quad\quad 
\Gamma^{\phi}{}_{\nu\rho} =
{\renewcommand{\arraystretch}{1.5}
\begin{pmatrix}
0 & 0 & 0 & ~~\mathcal{K}_{3}~~ \\
0 & 0 & 0 & ~~\frac{1}{r}~~ \\
0 & 0 &  0 & ~~\cot\theta~~ \\
~~\mathcal{K}_{3}~~ & ~~\frac{1}{r}~~ &  ~~\cot\theta~~ & 0
\end{pmatrix}}~,
\end{eqnarray}
and are governed by the time-dependent functions $\mathcal{K}_{1}$, $\mathcal{K}_{2}$ and $\mathcal{K}_{3}$, which are constrained by the curvature-free condition 
\begin{eqnarray*}
    \mathcal{K}_{3}(\mathcal{K}_{1}-\mathcal{K}_{3})-\dot{\mathcal{K}_{3}} &=& 0~,\\
    \mathcal{K}_{2}(\mathcal{K}_{1}-\mathcal{K}_{3})+\dot{\mathcal{K}_{2}} &=& 0~,\\
    \kappa + \mathcal{K}_{2}\mathcal{K}_{3} &=& 0.
\end{eqnarray*}
To derive the nonmetricity scalar $Q$, we make use of the definition given in Eqn. \eqref{NMS} and \eqref{CNS}. This leads to the expression for $Q$ as,
\begin{equation}\label{eq10}
    Q = -6H^{2}+9H\mathcal{K}_{3}+3\mathcal{K}_{3}\left(\mathcal{K}_{1}-\mathcal{K}_{3}\right) +\frac{3\mathcal{K}_{2}H}{a^{2}}-\frac{3\mathcal{K}_{2}\left(\mathcal{K}_{1}+\mathcal{K}_{3}\right)}{a^{2}}~.
\end{equation}
If we consider, $\mathcal{K}_{1} = \gamma(t), \quad \mathcal{K}_{2} = \mathcal{K}_{3} = 0$, this case is referred to as \textbf{Connection $I$}. The corresponding nonmetricity scalar $Q$ and Friedmann equations are given as follows:
\begin{eqnarray}
    && Q = -6H^{2}~, \label{eq11}\\
    && 3H^{2}f_{Q}+\frac{1}{2}(f-Qf_{Q}) = \rho~,\label{eq12}\\
    && -2\frac{d(f_{Q}H)}{dt}-3H^{2}f_{Q}-\frac{1}{2}(f-Qf_{Q}) = p~.\label{eq13}
\end{eqnarray}
For the \textbf{Connection $II$}, the functions take the form $\mathcal{K}_{1} = \gamma(t) + \frac{\dot{\gamma}(t)}{\gamma(t)}, \quad \mathcal{K}_{2} = 0, \quad \mathcal{K}_{3} = \gamma(t)$ and the nonmetricity scalar $Q$, Friedmann equations are,
\begin{eqnarray}
    && Q = -6H^{2}+9\gamma H+3\dot{\gamma}~, \label{eq14}\\
    && 3H^{2}f_{Q}+\frac{1}{2}(f-Qf_{Q})+\frac{3\gamma}{2}\dot{Q}f_{QQ} = \rho~,\label{eq15}\\
    && -2\frac{d(f_{Q}H)}{dt}-3H^{2}f_{Q}-\frac{1}{2}(f-Qf_{Q})+\frac{3\gamma}{2}\dot{Q}f_{QQ} = p~,\nonumber\\
    \label{eq16}
\end{eqnarray}
whereas the nonmetricity scalar $Q$ and the corresponding Friedmann equations for \textbf{Connection $III$}, with parameterization $\mathcal{K}_{1} = -\frac{\dot{\gamma}(t)}{\gamma(t)}, \quad \mathcal{K}_{2} = \gamma(t), \quad \mathcal{K}_{3} = 0$ are given by,
\begin{eqnarray}
    && Q = -6H^{2}+\frac{3\gamma H}{a^{2}}+\frac{3\dot{\gamma}}{a^{2}}, \label{eq17}\\
    && 3H^{2}f_{Q}+\frac{1}{2}(f-Qf_{Q})-\frac{3\gamma}{2a^{2}}\dot{Q}f_{QQ} = \rho~,\label{eq18}\\
    && -2\frac{d(f_{Q}H)}{dt}-3H^{2}f_{Q}-\frac{1}{2}(f-Qf_{Q})+\frac{\gamma}{2a^{2}}\dot{Q}f_{QQ} = p~.\nonumber\\
    \label{eq19}
\end{eqnarray}
As observed, the nonmetricity scalar $Q$ and the Friedmann equations \eqref{eq11}–\eqref{eq13} corresponding to Connection $I$ do not depend on the function $\gamma(t)$, which is in accordance with the results obtained in the coincident gauge $\left(\Gamma^{\lambda}_{~~\mu\nu}=0\right)$ in the flat FLRW metric.

\section{Dynamical System Formalism}\label{DSA_Section}
\noindent Connection $I$ corresponds to coincident gauge choices and leads to a simplified dynamical formulation. In contrast, Connections $II$ and $III$ exhibit a similar structural form but differ in the behavior of the function $\gamma$, which introduces an additional degree of freedom \cite{Guzman_2024_110_124013}. When the choice $\gamma(t) = H(t)$ is imposed, the behavior of the nonmetricity scalar differs significantly between Connection $II$ and Connection $III$. For Connection $II$, this substitution leads to $Q = 3H^{2} + 3\dot{H}$, where the additional contributions remain unsuppressed throughout the cosmic evolution, allowing persistent deviation from the standard form. In contrast, for Connection $III$ one obtains $Q = -6H^2 + \frac{3H^2}{a^2} + \frac{3\dot{H}}{a^2}$, where the extra terms are scaled by $\frac{1}{a^2}$. As the Universe expands, these contributions decay rapidly and the model naturally approaches the standard behavior $Q \approx -6H^{2}$ at late times. Thus, while Connection $II$ with $\gamma(t) = H(t)$ retains stronger dynamical effects and the Friedmann equations becomes,
\begin{eqnarray}
&& 3H^{2}f_{Q}+\frac{1}{2}(f-Qf_{Q})+\frac{3H}{2}\dot{Q}f_{QQ} = \rho~,\nonumber\\
    && -2\frac{d(f_{Q}H)}{dt}-3H^{2}f_{Q}-\frac{1}{2}(f-Qf_{Q})+\frac{3H}{2}\dot{Q}f_{QQ} = p~.\label{eq19}
\end{eqnarray}
\noindent To construct the autonomous dynamical system in this formalism, the set of dimensionless variables can be defined as follows. Here we refer to the methodology presented in \cite{Santos2018}.
\begin{equation}
\begin{aligned}
x_1 &= \frac{\dot{f}_Q}{H\,f_Q}\,, \quad
x_2 &= \frac{f}{6\,H^2\,f_Q}\,, \quad
x_3 &= \frac{Q}{6\,H^2}\,, \quad
x_4 &= \frac{\rho_r}{3\,H^2\,f_Q}\,, \quad
x_5 &= \frac{\rho_m}{3\,H^2\,f_Q}~.
\end{aligned}
\end{equation}
 Here, an overdot represents the derivative with respect to cosmic time $t$. The standard density parameters can be calculated using the formula listed below,
 \begin{equation}
\Omega_m = x_5, \quad \Omega_r = x_4, \quad \Omega_{DE} = x_3 - x_2 - \frac{x_1}{2}\,.\label{Eq:density_parameters}
\end{equation}
As a part of construction, we require an additional variable $b$, \begin{equation}
b = \frac{Q f_{QQ}}{f_Q}\,.
\end{equation}
The field equation presented in \eqref{eq19} and the density parameters in Eq. \eqref{Eq:density_parameters} produce the constraint equation,
\begin{equation}
-\frac{x_1}{2} - x_2 + x_3 + x_4 + x_5 = 1 \,.\label{constrain_eq}
\end{equation}
 The general autonomous dynamical system is then obtained by differentiating the dynamical variables with respect to $N=ln(a)$ as,
 
\begin{align}
\frac{dx_1}{dN} &= -3x_1 - x_1 \frac{\dot{H}}{H^2} - x_1^2 \,, \nonumber\\[6pt]
\frac{dx_2}{dN} &= \frac{x_1 x_3}{b} - 2 \frac{\dot{H}}{H^2} x_2 - x_1 x_2 \,,\nonumber\\[6pt]
\frac{dx_3}{dN} &= \frac{x_1 x_3}{b} - 2 \frac{\dot{H}}{H^2} x_3 \,,\nonumber\\[6pt]
\frac{dx_4}{dN} &= -4x_4 - 2 \frac{\dot{H}}{H^2} x_4 - x_1 x_4 \,,\nonumber\\[6pt]
\frac{dx_5}{dN} &= -3x_5 - 2 \frac{\dot{H}}{H^2} x_5 - x_1 x_5 \,.\label{DSA_1}
\end{align}
Where,
\begin{equation}
\frac{\dot{H}}{H^2} = \frac{3}{2} \left[
- \frac{x_4}{3} + x_3 - \frac{x_1}{6} - x_2 - 1
\right]\,.
\end{equation}
To convert the system into an autonomous form, it is essential to determine the value of $b$. Therefore, we need to consider the various forms of $f(Q)$. Below, we discuss the dynamics of the Universe at different phases using two well-known models from the literature in detail.

\subsection{Power-law model}\label{power_law}

The power-law model $f(Q)=\alpha Q^{n}$ \cite{Jimenez_2020_101_103507,Lazkoz:2019sjl, Ayuso_2021_103_063505, Paliathanasis2023} is adopted because $\alpha=n=1$, it reduces to GR, it provides the simplest nonlinear generalization of GR that models can describe the late-time Universe acceleration and it is also compatible with BBN constraints \cite{Anagnostopoulos_2023_83_58}. This form naturally admits a scaling solution and a clear dynamical structure, allowing the description of radiation and matter-like and accelerated phases within a single framework. With only one parameter $n$, it enables a controlled study of deviations from GR while making it easier to analyze the impact of connection-dependent effects on the expansion history of the Universe \cite{Khyllep_2023_107_044022}. The partial derivatives of the model are given by $f_Q = \alpha n Q^{n-1}$ and $f_{QQ} = \alpha n (n-1) Q^{n-2}$. These expressions lead to a dependency relation between the variables $x_3$ and $x_2$, namely $x_2 = \frac{x_3}{n}$. Using this relation, the general dynamical system in Eq. \eqref{DSA_1} reduces to the following form:

\begin{align}
\frac{dx_1}{dN} 
&= \frac{x_1 \Big(-3 n (x_1 + 2x_3 + 2) + 2 n x_4 + 6 x_3 \Big)}{4n}\,,
\nonumber \\[8pt]
\frac{dx_3}{dN} 
&= \frac{1}{2} x_3 \Bigg(
\frac{2 x_1}{n-1} + \frac{6 x_3}{n} + x_1 - 6 x_3 + 2 x_4 + 6
\Bigg)\,,
\nonumber \\[8pt]
\frac{dx_4}{dN} 
&= x_4 \Bigg(
3 \left(\frac{1}{n} - 1\right) x_3 - \frac{x_1}{2} + x_4 - 1\Bigg)\,,
\nonumber \\[8pt]
\frac{dx_5}{dN} 
&= x_5 \Bigg(
3 \left(\frac{1}{n} - 1\right) x_3 + x_4
\Bigg) - \frac{x_1 x_5}{2}\,.
\label{DSA_2}
\end{align}
To analyze the cosmological model presented in this system, we derive the critical points by setting the derivatives to zero: $\frac{dx_i}{dN}=0\,, i=1,2,3,4$. The critical points, along with the corresponding eigenvalues of the Jacobian matrix for our system, are detailed in Table \ref{model1_combined} on the left. We employed the eigenvalue method to assess the stability of these critical points. Additionally, we calculated significant cosmological parameters, such as the deceleration parameter and the effective EoS \(\omega_{tot}\), to identify which critical point corresponds to a specific cosmological epoch. The values of the standard density parameter at each critical point are summarized in Table \ref{model1_combined} on the right side. The detailed analysis of each critical point is presented separately as follows:
\begin{table*}[!htb]
\centering
{\small

\setlength{\tabcolsep}{4pt} 
\renewcommand{\arraystretch}{1.2} 

\begin{minipage}{0.48\textwidth}
\centering
\begin{tabular}{|c|c|c|c|c|c|}
\hline
\textbf{Critical Points} 
& \textbf{$x_1$}
& \textbf{$x_3$} 
& \textbf{$x_4$} 
& \textbf{$x_5$} 
& \textbf{Eigenvalues} \\
\hline \hline
$P_1$ & $0$ & $0$ & $0$ & $x_5$ & $\left\{0,-\frac{3}{2},-1,3\right\}$ \\
\hline
$P_2$ & $0$ & $\frac{n}{n-1}$ & $0$ & $0$ & $\{-4,-3,-3,-3\}$ \\
\hline
$P_3$ & $-2$ & $0$ & $0$ & $0$ & $\left\{0,\frac{2 (n-2)}{n-1},1,\frac{3}{2}\right\}$ \\
\hline
$P_4$ & $0$ & $0$ & $1$ & $0$ & $\{-1,1,1,4\}$ \\
\hline
\end{tabular}
\caption*{(a) Critical Points}
\end{minipage}
\hfill
\begin{minipage}{0.48\textwidth}
\centering
\begin{tabular}{|c|c|c|c|c|c|c|}
\hline
\textbf{C.P.}  & \textbf{Stability}  & \textbf{$q$}  & \boldmath{$\omega_{\text{tot}}$} & \boldmath{$\Omega_r$}  & \boldmath{$\Omega_m$}  & \boldmath{$\Omega_{DE}$} \\
\hline \hline
$P_1$  & Saddle & $\frac{1}{2}$  & $0$ & $0$  & $1$  & $0$ \\
\hline
$P_2$  & Stable & $-1$  & $-1$  & $0$  & $0$  & $1$ \\
\hline
$P_3$  & \begin{tabular}{@{}c@{}}
Saddle at $1<n<2$ \\
Unstable at $n<1 \lor n>2$
\end{tabular} & $0$  & $-\frac{1}{3}$  & $0$  & $0$ & $1$ \\
\hline
$P_4$  & Saddle & $1$ & $\frac{1}{3}$ & $1$ & $0$ & $0$ \\
\hline
\end{tabular}
\caption*{(b) Stability and Cosmological Parameters}
\end{minipage}

\caption{Critical points for the power-law model along with their stability, deceleration parameter, EoS parameter and density parameters.}
\label{model1_combined}
}
\end{table*}

\begin{figure}[ht]
    \centering
    
    \fbox{%
    \begin{minipage}{0.95\linewidth}
        \centering
        
        \begin{subfigure}[b]{0.30\linewidth}
            \centering
            \includegraphics[width=\linewidth]{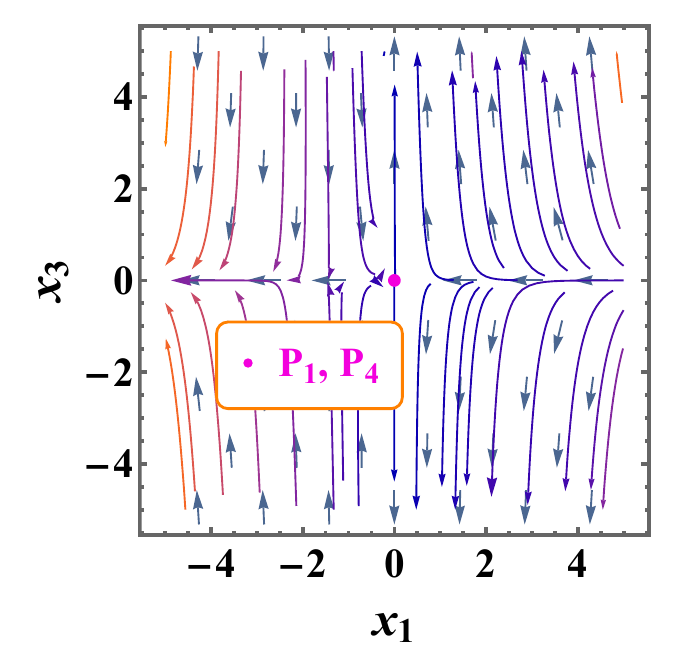}
            \caption{Phase portrait with critical points $P_1, P_4$.}
         \label{fig:phase1}
        \end{subfigure}
        \hfill
        \begin{subfigure}[b]{0.30\linewidth}
            \centering
            \includegraphics[width=\linewidth]{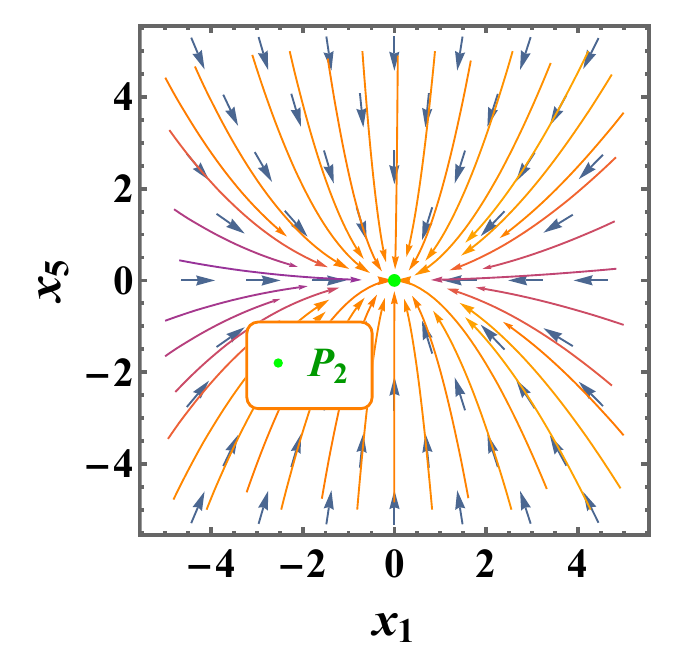}
            \caption{Phase portrait with critical point $P_2$.}
            \label{fig:phase2}
        \end{subfigure}
        \hfill
        \begin{subfigure}[b]{0.30\linewidth}
            \centering
            \includegraphics[width=\linewidth]{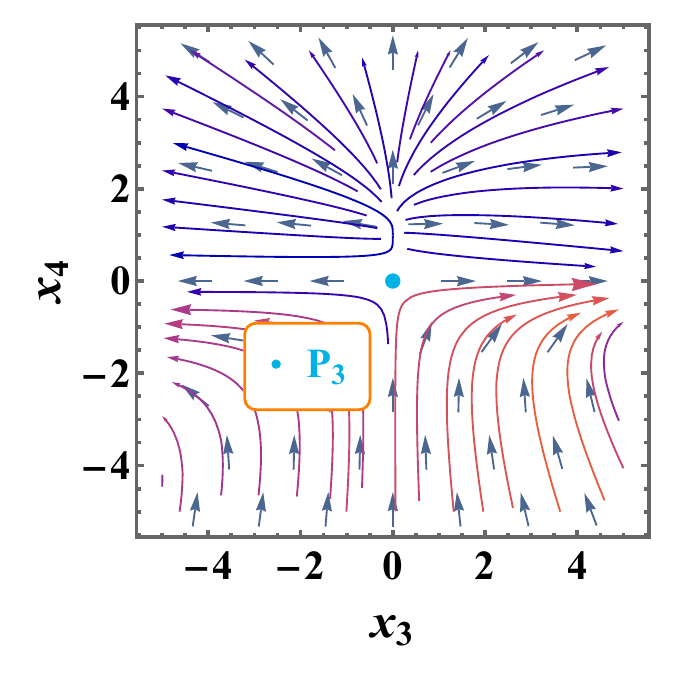}
            \caption{Phase portrait with critical point $P_3$.}
            \label{fig:phase3}
        \end{subfigure}
        
    \end{minipage}}
    
    \caption{Phase portraits of the system illustrating the critical points $P_1$ to $P_5$.}
    \label{fig:phase_all_m1}
\end{figure}

\begin{itemize}
\item{\bf Critical Point $P_1$:} The critical point $P_1$ is classified as a matter-dominated critical point with an EoS and deceleration parameter $\omega_{tot} = 0$. This point represents the standard scenario for a matter-dominated Universe where $\Omega_{m} = 1$, meaning there is no contribution from radiation or other components. At this critical point, the eigenvalues exhibit both positive and negative values, which indicates that the point behaves like a saddle. This saddle nature is further illustrated by the phase space trajectories, which, as depicted in Fig. \ref{fig:phase1}, are observed diverging from the critical point, thereby confirming its saddle characteristic.

\item{\bf Critical Point $P_2$:} This significant critical point is characterized as a DE-dominated state, where both the deceleration parameter and the EoS total are equal to $-1$. This condition illustrates the widely recognized phenomenon of cosmic acceleration in the late Universe, primarily driven by dark energy, as indicated by $\Omega_{DE}=1$. Furthermore, all eigenvalues associated with this critical point exhibit negative values, confirming its stability. The attracting characteristics of this critical point can be further explored in Fig. \ref{fig:phase2}.

\item{\bf Critical Point $P_3$:}
This critical point signifies a significant transition in the Universe, shifting from a phase of deceleration to acceleration. At this juncture, the deceleration parameter is $q=0$ and the EoS parameter is $\omega_{tot}=-\frac{1}{3}$. It marks the moment when the dominance in the Universe shifts from dark matter to dark energy, characterized by $\Omega_{DE}=1$. The presence of a zero eigenvalue at this point suggests that it is non-hyperbolic, meaning its stability must be assessed by examining the signs of the remaining eigenvalues. Within the parameter range of $1<n<2$, this critical point behaves like a saddle point, while it exhibits instability for values of $n$ such that $1<n$ or $n>2$. The saddle characteristics of this critical point can be visually represented through phase portraits, as shown in Fig. \ref{fig:phase3}.

\item{\bf Critical Point $P_4$:} This critical point describes a significant early decelerating phase in the evolution of the Universe when radiation played a dominant role. At this point, we observe a deceleration parameter $q = 1$ and an EoS  $\omega_{tot} = \frac{1}{3}$. These findings illustrate how the model successfully describes all major phases of the development of the Universe. The eigenvalues corresponding to this critical point show both positive and negative signs, indicating that it behaves as a saddle point. This behavior has also been corroborated by the phase space analysis depicted in Fig. \ref{fig:phase3}.
\end{itemize}

\begin{figure}[ht]
    \centering
    
    \fbox{%
    \begin{minipage}{0.98\linewidth}
    \centering
    
    \begin{subfigure}[b]{0.32\linewidth}
        \centering
        \includegraphics[width=\linewidth]{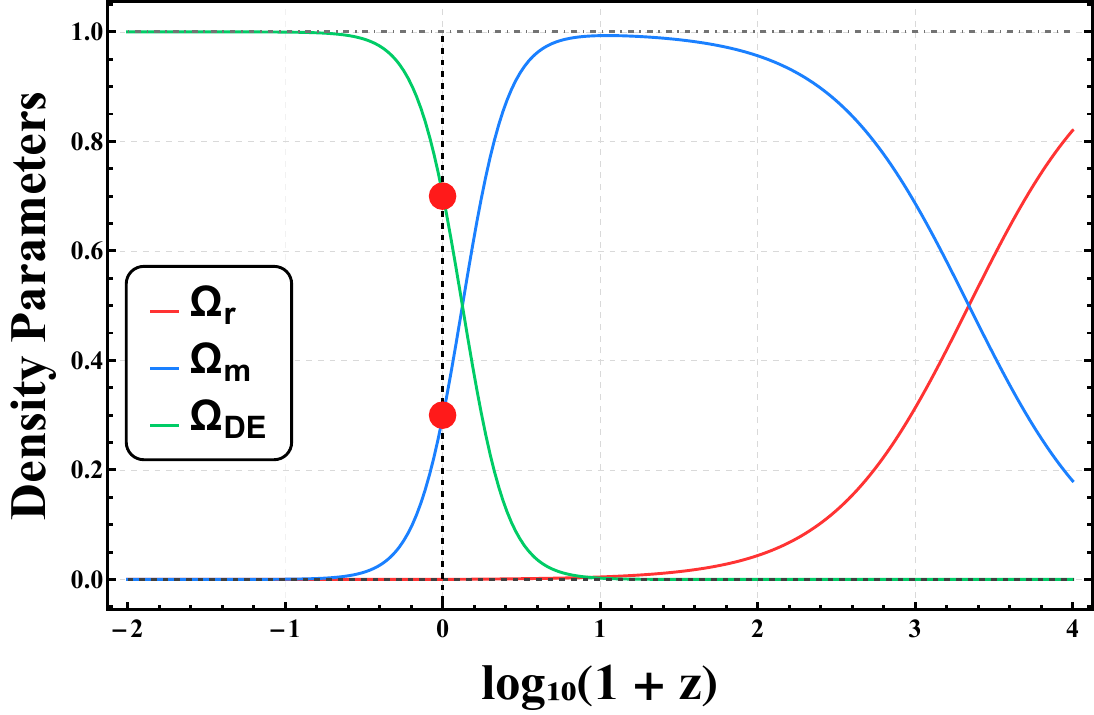}
        \caption{Evolution of the density parameters.}
        \label{fig:density_parameter_m_1}
    \end{subfigure}
    \hfill
    \begin{subfigure}[b]{0.32\linewidth}
        \centering
        \includegraphics[width=\linewidth]{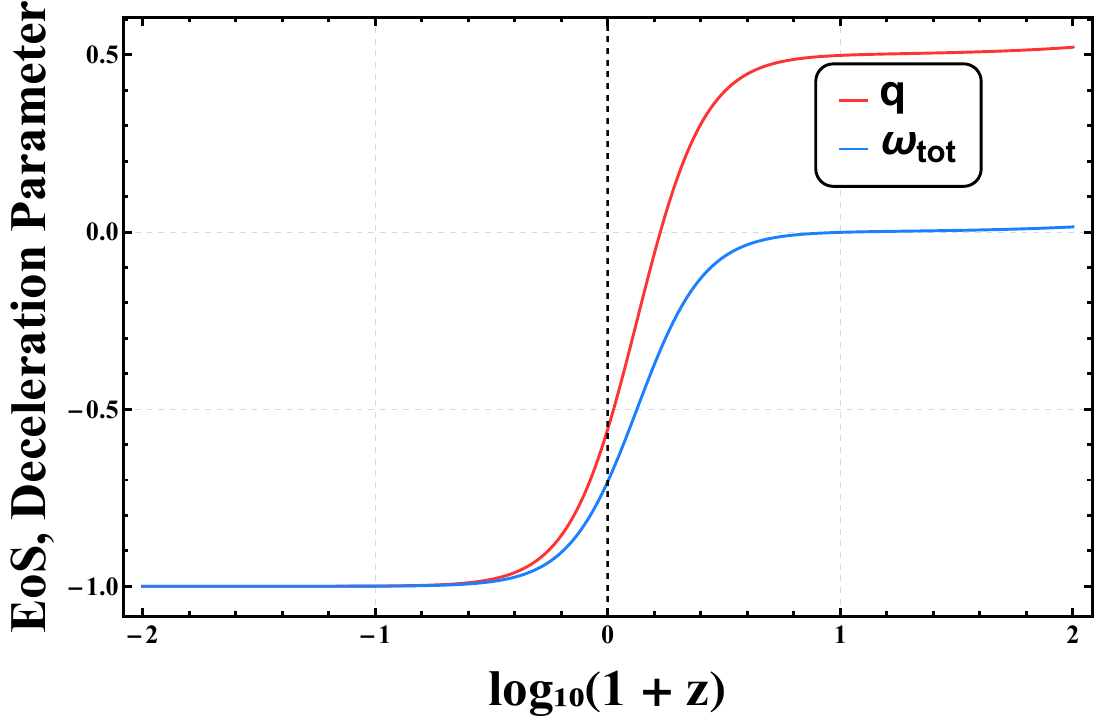}
        \caption{Evolution of $q, \omega_{tot}$.}
        \label{fig:deceleration_parameter_m_1}
    \end{subfigure}
    \hfill
    \begin{subfigure}[b]{0.32\linewidth}
        \centering
        \includegraphics[width=\linewidth]{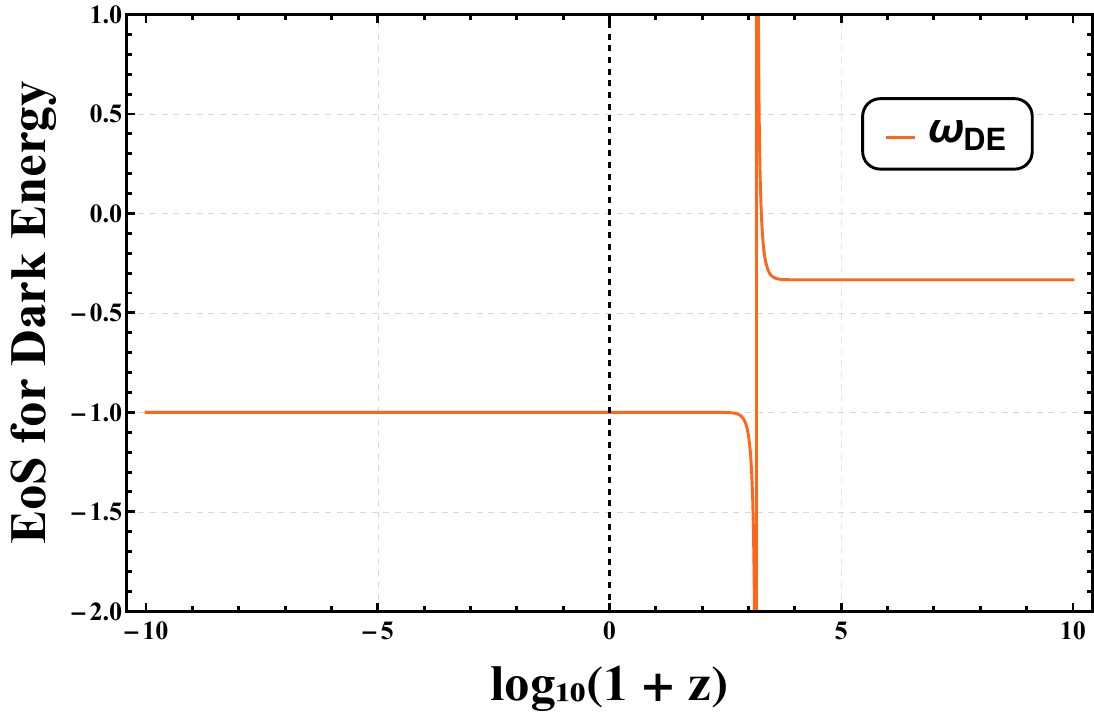}
        \caption{EoS parameter $\omega_{DE}$.}
        \label{fig:eos_parameter_m_1}
    \end{subfigure}
    
    \end{minipage}
    }
    
    \caption{Cosmological evolution with respect to redshift: (a) density parameter $\Omega$, (b) deceleration parameter $q$ and (c) EoS parameter $\omega_{DE}$ with the initial conditions $n=2.1, x_{1c}=10^{-10}, x_{3c}=10^{-7}, x_{4c}=10^{-0.9}, x_{5c}=10^{-0.4}$.}
    \label{fig:combined_m_1}
\end{figure}
\noindent To gain a clearer understanding of the dynamics, we illustrate the behavior of the standard density parameters associated with radiation, matter and DE. Additionally, we provide plots for the EoS for both the total energy and DE, as well as the important deceleration parameter, depicted in Fig. \ref{fig:combined_m_1} and \ref{fig:combined_m_2} for Model I and Model II, respectively. In Fig. \ref{fig:combined_m_1}, we observe that the standard density parameters follow a sequence of dominance, beginning with radiation, transitioning to matter and ultimately showing DE predominance in the present and future. Currently, the matter contribution is approximately 0.3, while dark energy constitutes about 0.7 inclined with \cite{Gonzalez-Espinoza:2020jss}, as highlighted by the dark red dots. Both the deceleration parameter and the EoS indicate a shift from early deceleration to late-time acceleration, with values of $q_0 \approx -0.56$ $\omega_0 \approx -0.70$ which are compatible with ~\cite{Planck:2018vyg, Capozziello:2014}.

\subsection{Logarithmic model}
\label{Logarithmic_law}

The second model considered in this study is the logarithmic form of $f(Q)$. This model has been previously explored for its consistency with Big Bang Nucleosynthesis (BBN) constraints and for its dynamical behavior \cite{ANAGNOSTOPOULOS20,Vishwakarma2023}. However, those analyses were carried out without focusing on a specific class of connection and therefore, the role of connection-dependent effects in shaping the cosmological evolution remains to be examined in detail. Using Eq. \eqref{constrain_eq} and $b = -1$ for the logarithmic model, the dynamical system presented in Eq. \eqref{DSA_1} reduces to a four-dimensional system, taking the following form:

\begin{align}
\frac{dx_1}{dN} &= \frac{1}{2} x_1 \left(-3x_1 + 4x_4 + 3x_5 - 6\right), \nonumber\\
\frac{dx_3}{dN} &= x_3 \left(-2x_1 + 4x_4 + 3x_5\right), \nonumber\\
\frac{dx_4}{dN} &= x_4 \left(-2x_1 + 4x_4 + 3x_5 - 4\right), \nonumber\\
\frac{dx_5}{dN} &= x_5 \left(-2x_1 + 4x_4 + 3x_5 - 3\right).\label{DS_3}
\end{align}
The critical points along with their corresponding eigenvalues are summarized in Table \ref{model2_combined} on the left. On the right, you will find detailed information regarding various cosmological parameters. A thorough analysis of each specific critical point is provided in the following sections.

\begin{table*}[!htb]
\centering
{\small

\begin{minipage}{0.48\textwidth}
\centering
\addtolength{\tabcolsep}{-2pt}
\begin{tabular}{|c|c|c|c|c|c|}
\hline
\textbf{Critical Points} 
& \textbf{$x_1$}
& \textbf{$x_3$} 
& \textbf{$x_4$} 
& \textbf{$x_5$} 
& \textbf{Eigenvalues} \\
\hline \hline
$C_1$ & $0$ & $0$ & $0$ & $1$ & $\left\{3,3,-\frac{3}{2},-1\right\}$ \\
\hline
$C_2$ & $0$ & $x_3$ & $0$ & $0$ & $\{-4,-3,-3,0\}$ \\
\hline
$C_3$ & $-2$ & $0$ & $0$ & $0$ & $\{4,3,1,0\}$ \\
\hline
$C_4$ & $0$ & $0$ & $1$ & $0$ & $\{4,4,-1,1\}$ \\
\hline
\end{tabular}
\caption*{(a) Critical Points}
\end{minipage}
\hfill
\begin{minipage}{0.48\textwidth}
\centering
\addtolength{\tabcolsep}{-2pt}
\begin{tabular}{|c|c|c|c|c|c|c|}
\hline
\textbf{C.P.}  & \textbf{Stability}  & \textbf{$q$}  & \boldmath{$\omega_{\text{tot}}$} & \boldmath{$\Omega_r$}  & \boldmath{$\Omega_m$}  & \boldmath{$\Omega_{DE}$} \\
\hline \hline
$C_1$  & Saddle & $\frac{1}{2}$  & $0$ & $0$  & $1$  & $0$ \\
\hline
$C_2$  & Non-hyperbolic Stable & $-1$  & $-1$  & $0$  & $0$  & $1$ \\
\hline
$C_3$  & Unstable & $0$  & $-\frac{1}{3}$  & $0$  & $0$ & $1$\\
\hline
$C_4$  & Saddle & $1$ & $\frac{1}{3}$ & $1$ & $0$ & $0$ \\
\hline
\end{tabular}
\caption*{(b) Stability and Cosmological Parameters}
\end{minipage}

\caption{Critical points for the logarithmic model along with their stability, deceleration parameter, EoS parameter and density parameters.}
\label{model2_combined}
}
\end{table*}

\begin{figure}[ht]
    \centering
    
    \fbox{%
    \begin{minipage}{0.95\linewidth}
        \centering
        
        \begin{subfigure}[b]{0.30\linewidth}
            \centering
            \includegraphics[width=\linewidth]{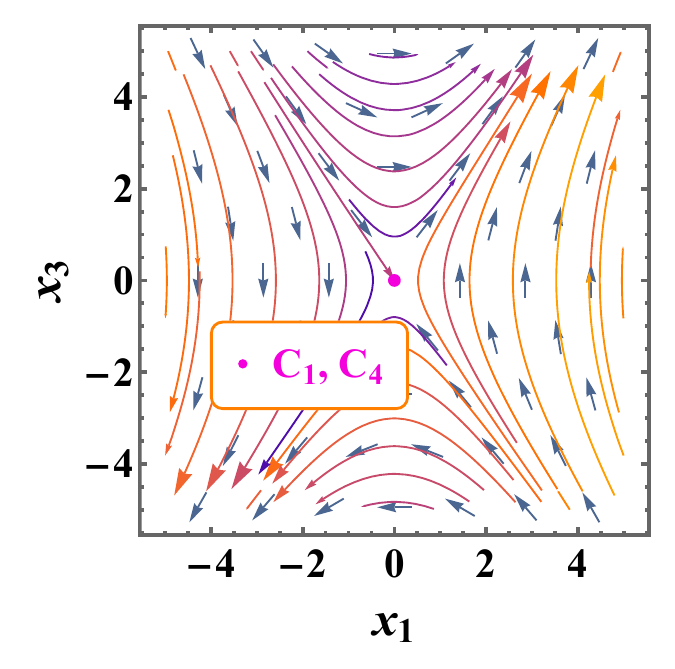}
            \caption{Phase portrait with critical points $C_1, C_4$.}
            \label{fig:phase1m2}
        \end{subfigure}
        \hfill
        \begin{subfigure}[b]{0.30\linewidth}
            \centering
            \includegraphics[width=\linewidth]{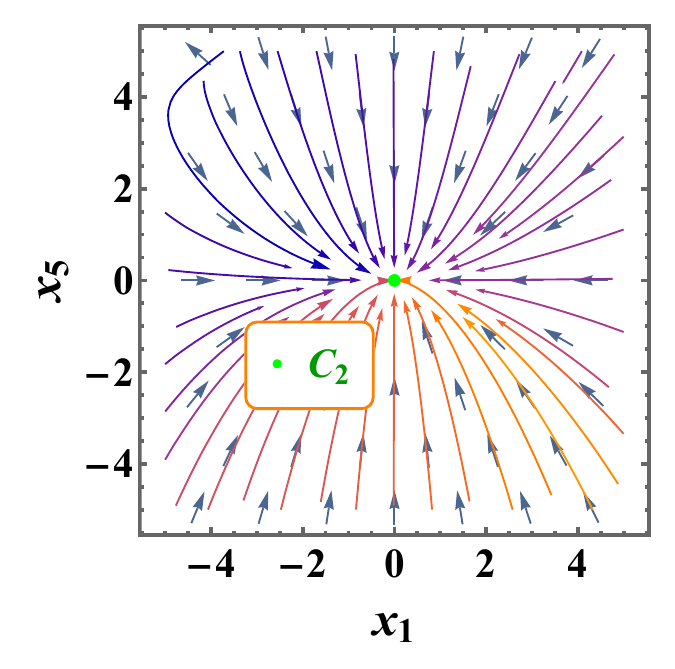}
            \caption{Phase portrait with critical point $C_2$.}
            \label{fig:phase2m2}
        \end{subfigure}
        \hfill
        \begin{subfigure}[b]{0.30\linewidth}
            \centering
            \includegraphics[width=\linewidth]{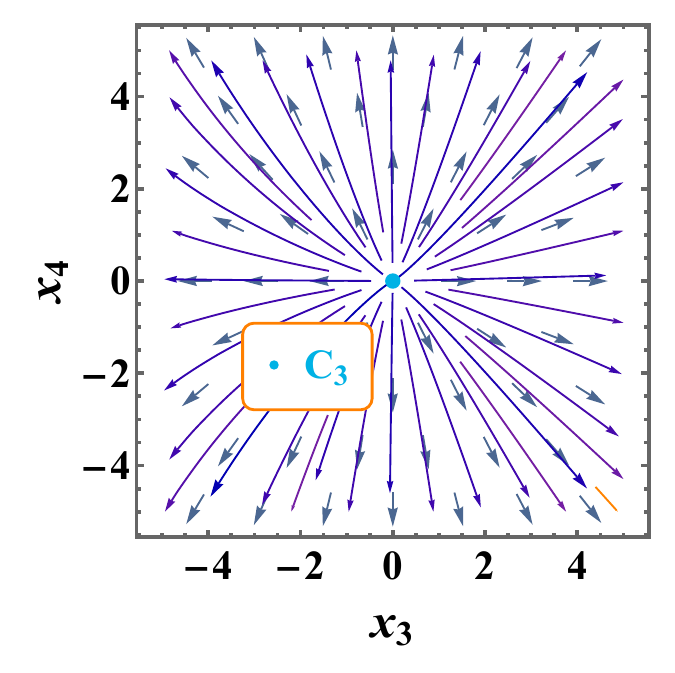}
            \caption{Phase portrait with critical point $C_3$.}
            \label{fig:phase3m2}
        \end{subfigure}
        
    \end{minipage}}
    
    \caption{Phase portraits of the system illustrating the critical points $C_1$ to $C_5$.}
    \label{fig:phase_all}
\end{figure}

\begin{itemize}
    \item{\bf Critical Point $C_1$:} The corresponding EoS parameter and deceleration parameter are given by $\omega_{\text{tot}} = 0$ and $q = \frac{1}{2}$, respectively. This behavior of the critical point characterizes a matter-dominated, decelerating phase of the Universe. The density parameters take the values $\Omega_{r} = 0$, $\Omega_{m} = 1$, and $\Omega_{DE} = 0$, confirming that this point represents a standard matter-dominated epoch (see Table \ref{model2_combined}). Furthermore, the presence of both positive and negative eigenvalues in the Jacobian matrix indicates that the critical point is an unstable saddle point, which is also evident from Fig. \eqref{fig:phase1m2}.
    \item{\bf Critical Point $C_2$:} At this critical point, the density parameters take the values $\Omega_{r}=0$, $\Omega_{m}=0$, and $\Omega_{DE}=1$, indicating a dark energy–dominated phase of the Universe. The accelerated nature of this phase is supported by the corresponding values of the EoS parameter, $\omega_{\text{tot}}=-1$ and the deceleration parameter, $q=-1$, as listed in Table \ref{model2_combined}. This configuration corresponds to a de-Sitter solution. The Jacobian matrix evaluated at this critical point possesses eigenvalues with negative real parts along with a zero eigenvalue, implying that the point is non-hyperbolic \cite{aulbach1984continuous, Coley:1999uh}. Therefore, the stability analysis of this point requires the application of Center Manifold Theory, which will be carried out in the following subsection. Moreover, this critical point acts as a late-time attractor, and its attracting behavior in the phase space is illustrated in Fig. \eqref{fig:phase2m2}.
    \subsubsection*{Center Manifold Theory}
    To assess the stability of the critical point $C_{2}$, we employ Center Manifold Theory (CMT) for the dynamical system defined by Eqs. \eqref{DS_3}. The Jacobian matrix of the corresponding autonomous system \eqref{DS_3}, evaluated at the critical point $C_{2}$, is given by
    \begin{equation*}
    J(C_{2}) = 
    \begin{bmatrix}
    -3 & 0  & 0  & 0 \\
    -2x_{3} & 0 & 4x_{3} & 3x_{3} \\
    0 & 0 & -4 & 0  \\
    0 & 0 & 0 & -3  
    \end{bmatrix}
    \end{equation*}
    The eigenvalues of the system are $(-4, -3, -3, 0)$, with the corresponding eigenvectors given by $[0, -x_{3}, 1, 0]^{T}$, $\left[\frac{3}{2}, 0, 0, 1\right]^{T}$, $\left[\frac{3}{2x_{3}}, 1, 0, 0\right]^{T}$ and $[0, 1, 0, 0]^{T}$. To proceed further, the system is transformed into a standard form. In order to shift the critical point $C_{2} = (0, x_{3}, 0, 0)$ to the origin of the phase space, we introduce a new set of variables defined as $U = x_{1}$, $V = x_{3} - a$ (where $a \in \mathbb{R}$), $W = x_{4}$, and $R = x_{5}$. In terms of these new coordinates, the system of equations can be rewritten as follows:
    \begin{align}
\begin{pmatrix}
U'\\ 
V' \\ 
W'\\ 
R'
\end{pmatrix}= 
\begin{pmatrix}
-3 & 0 & 0 & 0\\
0 & 0 & 0 & 0 \\
0 & 0 & -4 & 0  \\
0 & 0 & 0 & -3 
\end{pmatrix} 
\begin{pmatrix}
U\\ 
 V\\ 
 W\\ 
 R    
\end{pmatrix}+\begin{pmatrix}
 non\\ linear\\ term   
\end{pmatrix}. 
\end{align}
In the standard formulation, the variables $U$, $W$ and $R$ are identified as stable variables, while $V$ acts as the central variable in the transformed system. According to CMT, the manifold can be described by a continuously differentiable function. In the present case, the center manifold is defined by the relations $U = h_{1}(V)$, $W = h_{3}(V)$ and $R = h_{4}(V)$.
\begin{eqnarray}
    U' = \frac{dh_{1}}{dV}V',\quad\quad    W' = \frac{dh_{3}}{dV}V',\quad\quad R' = \frac{dh_{4}}{dV}V'.
\end{eqnarray}
Utilizing the framework given in a quasi-linear partial differential equation, we now construct the corresponding approximation as,
\begin{eqnarray}\label{QS}
    \mathcal{N}_{1}(h_{1}(V)) =  \frac{dh_{1}}{dV}V' - U',\quad\quad
    \mathcal{N}_{1}(h_{3}(V)) =  \frac{dh_{3}}{dV}V' - W',\quad\quad
    \mathcal{N}_{1}(h_{4}(V)) =  \frac{dh_{4}}{dV}V' - R'.
\end{eqnarray}
After substituting the transformation into the system given in Eqn. \eqref{DS_3}, we obtain the following set of equations expressed in the new coordinate system:
\begin{align}
U' &= 2 U W-3 U+\frac{3 R U}{2}-\frac{3 U^2}{2}~, \nonumber\\
V' &= 3 a R-2 a U+4 a W+3 R V-2 U V+4 V W~, \nonumber\\
W' &= 3 R W-2 U W+4 W^2-4 W~, \nonumber\\
R' &= -3 R + 3 R^2 - 2 R U + 4 R W~.\label{DS_3}
\end{align}
For the zeroth-order approximation, we get vanishing manifold functions. Henceforth, we can derive a first-order approximation of the manifold function as, 
\begin{eqnarray}
 N_{1}(h_{1}(V)) &=& V \left(a_{1} b_{1}+2a_{1}b_{3}+\frac{3a_{1}b_{4}}{2}+3a_{1}-2a_{3}b_{1}-\frac{3a_{4}b_{1}}{2}\right)+\frac{3 b_{1}^2}{2}-2b_{1}b_{3}-\frac{3b_{1} b_{4}}{2}+3b_{1}~, \nonumber\\
 N_{1}(h_{3}(V)) &=& V (2a_{1}b_{3}-4a_{3}b_{3}+4 a_{3}-3 a_{4} b_{3})+2 b_{1} b_{3}-4b_{3}^2-3b_{3} b_{4}+4b_{3}~, \nonumber\\
 N_{1}(h_{4}(V)) &=& V (2a_{1}b_{4}-4a_{3}b_{4}-3a_{4}b_{4}+3a_{4})+2b_{1} b_{4}-4 b_{3} b_{4}-3 b_{4}^2+3b_{4}~.
\end{eqnarray}
In this case, the center manifold is given by the following expression,
\begin{equation}
    V'= V \left(-3 b_{1}^2+12b_{1} b_{3}+9b_{1} b_{4}-6 b_{1}-16b_{3}^2-24 b_{3} b_{4}+16b_{3}-9 b_{4}^2+9b_{4}\right) + higher~~order~~term
\end{equation}
By CMT, the critical point $C_{2}$ exhibits stable behavior for $V\neq0$ and \newline $\left((\sqrt{2}b_1-2\sqrt{2}b_3-\sqrt{2}b_4+\sqrt{2} )^2+(b_{1}-2b_{3}-b_{4}+1)^{2}+(4b_{3}^{2}+b_{4}(1+b_{1}+4b_{3})-4b_{3}-3)\right)>0$. The unstable region is shown in Fig. \ref{RS}.
\begin{figure}[ht]
    \centering
    \includegraphics[width=0.50\linewidth]{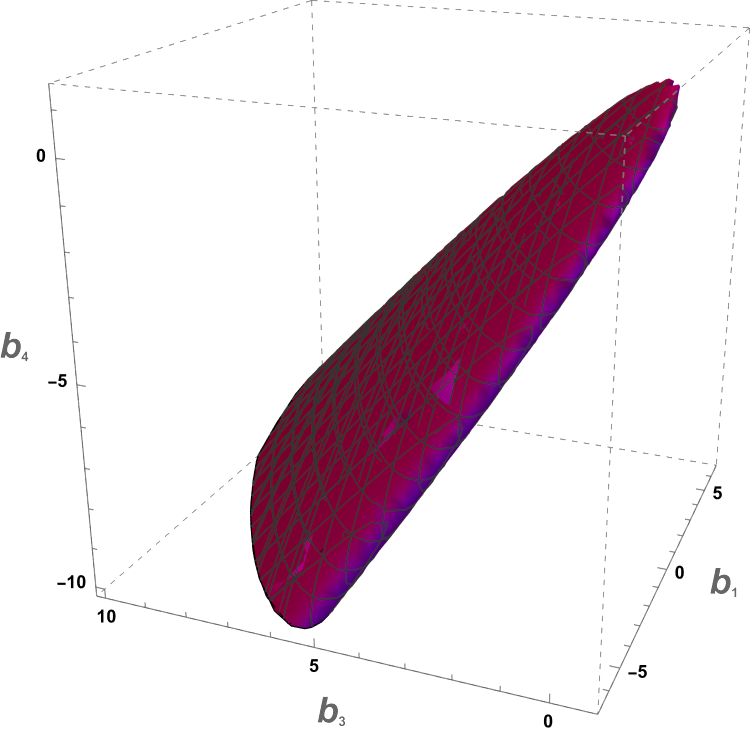}
    \caption{Unstable region for the critical point $C_{2}$ utilizing CMT.}
    \label{RS}
\end{figure}

    \item{\bf Critical Point $C_3$:} This critical point highlights the transition phase in the evolution of the Universe, where it shifts from a state of deceleration to acceleration, characterized by parameters $\omega_{tot}=\frac{-1}{3}$ and $q=0$. At this juncture, the dominance in the Universe transitions from matter to DE, with the density parameter $\Omega_{DE}=1$, as illustrated in Table \ref{model2_combined}. This point is non-hyperbolic due to vanishing eigenvalues, and its instability is evidenced by the positive signs of the remaining eigenvalues. Consequently, the trajectories in the phase space are diverging from this critical point, which is further illustrated in the phase space plot in Fig. \eqref{fig:phase3m2}.
    
    \item{\bf Critical Point $C_4$:} The values $q=1$, $\omega_{\text{tot}}=\frac{1}{3}$ indicate that this critical point represents an early decelerating, radiation-dominated phase of the Universe. As shown in Table \ref{model2_combined}, the presence of both positive and negative eigenvalues classifies this point as a saddle. The corresponding phase-space behavior, characteristic of saddle-type dynamics, is illustrated in Fig. \eqref{fig:phase1m2}. This analysis highlights the effectiveness of the Logarithmic model in examining the significant epochs of the evolution of the Universe.
\end{itemize}
\begin{figure}[H]
    \centering
    
    \fbox{%
    \begin{minipage}{0.98\linewidth}
    \centering
    
    \begin{subfigure}[b]{0.32\linewidth}
        \centering
        \includegraphics[width=\linewidth]{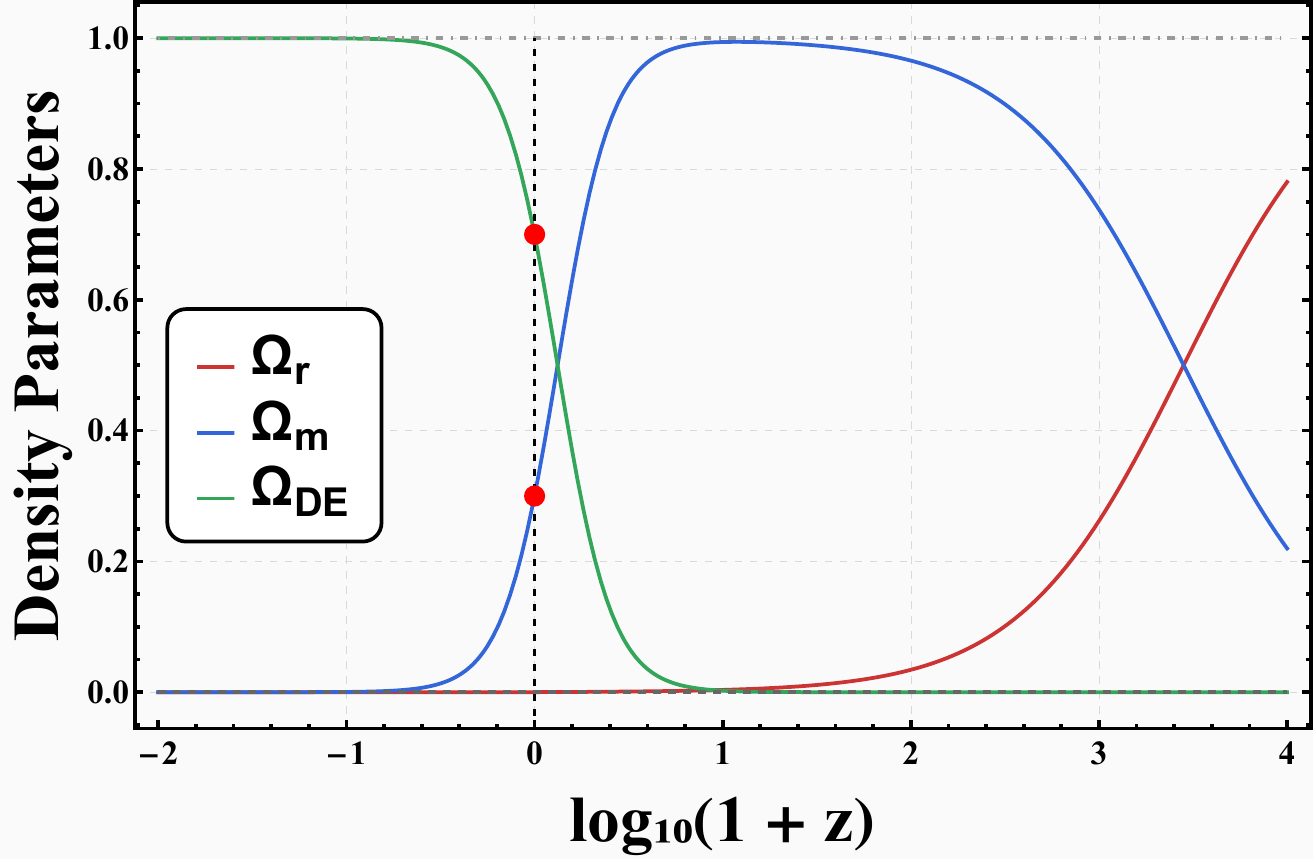}
        \caption{Evolution of the density parameters.}
        \label{fig:density_parameter_m_2}
    \end{subfigure}
    \hfill
    \begin{subfigure}[b]{0.32\linewidth}
        \centering
        \includegraphics[width=\linewidth]{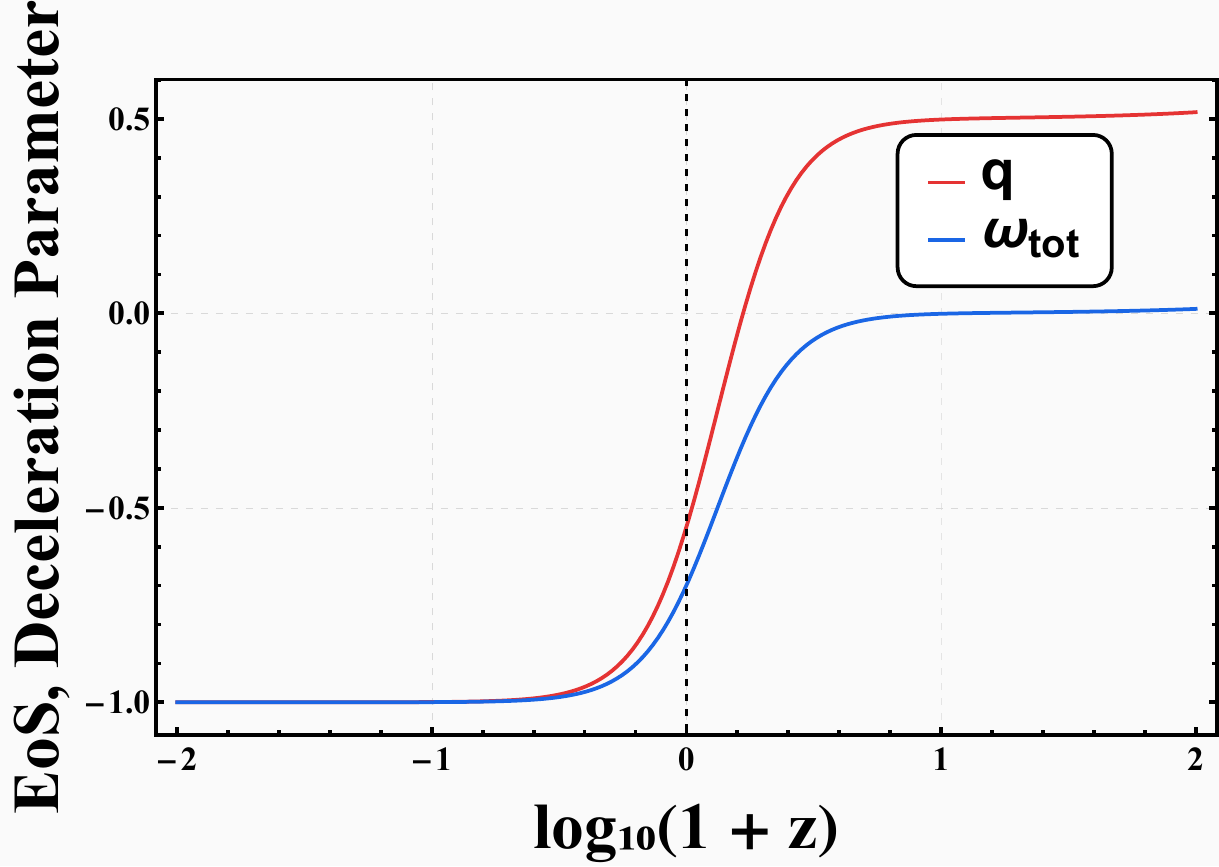}
        \caption{Evolution of $q, \omega_{tot}$.}
        \label{fig:deceleration_parameter_m_2}
    \end{subfigure}
    \hfill
    \begin{subfigure}[b]{0.32\linewidth}
        \centering
        \includegraphics[width=\linewidth]{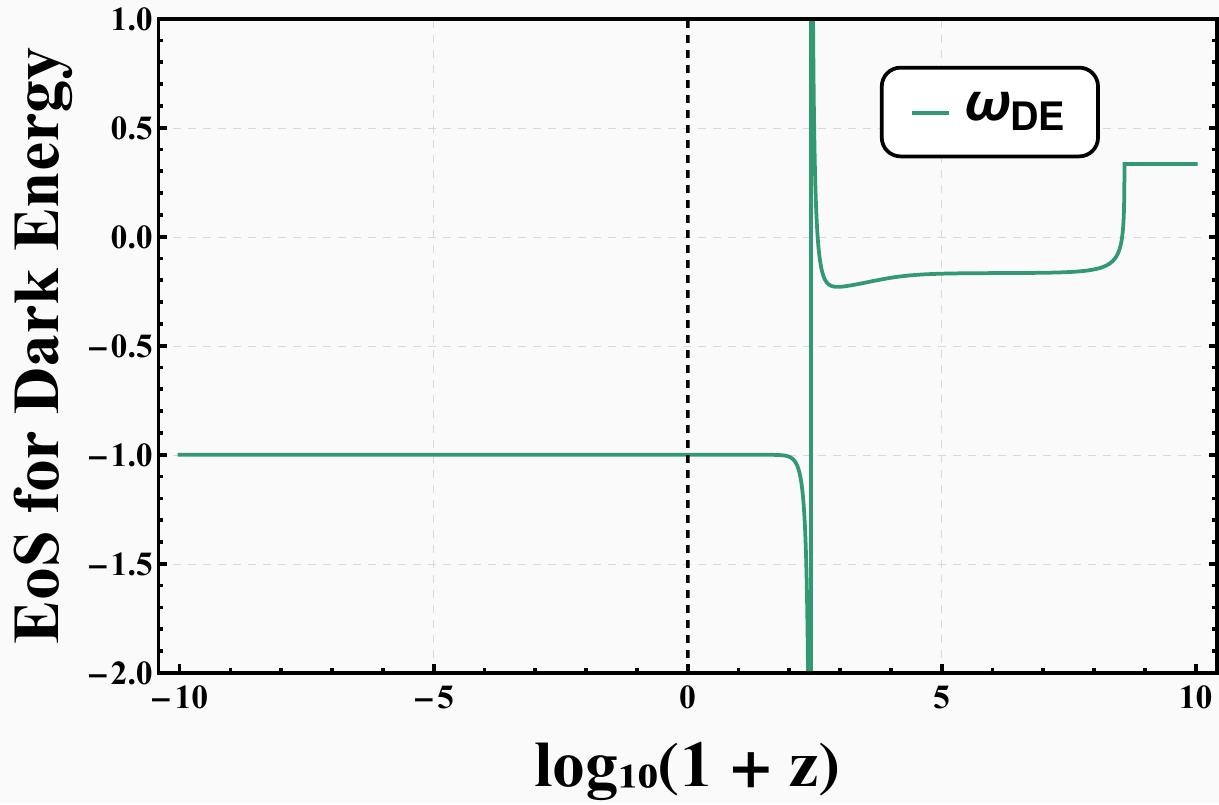}
        \caption{EoS parameter $\omega_{DE}$.}
        \label{fig:eos_parameter_m_2}
    \end{subfigure}
    
    \end{minipage}
    }
    
    \caption{Cosmological evolution with respect to redshift: (a) density parameter $\Omega$, (b) deceleration parameter $q$ and (c) EoS parameter $\omega_{DE}$ with the initial conditions $x_{1c}=10^{-10}, x_{3c}=10^{-32}, x_{4c}=10^{-3.45}, x_{5c}=10^{-0.2}$.}
    \label{fig:combined_m_2}
\end{figure}
\noindent In Figure \ref{fig:density_parameter_m_2}, we illustrate the behavior of the standard density parameter, which aligns well with key epochs in the evolution of the Universe and is consistent with current observations \cite{Gonzalez-Espinoza:2020jss}. The deceleration parameter and the total equation of state (EoS) parameter exhibit a transition from early deceleration to late-time acceleration, with specific values of $q_{0} = -0.558$ and $\omega_{0} = -0.699$. These results are in agreement with findings from ~\cite{Planck:2018vyg, Capozziello:2014}. Notably, one of the primary distinctions in the analyses of both models is the subtle variation in the behavior of $\omega_{de}$ during the early stages of the development of the Universe. As time progresses toward the present and into the future, the plot for $\omega_{de}$ approaches $-1$, suggesting that the model is consistent with the $\Lambda$CDM framework both at present and in the future.

\section{Final Remarks and Conclusion}\label{conclusion}
In this work, we have investigated the cosmological dynamics of covariant $f(Q)$ gravity, specifically using connection II within a dynamical system framework. The connection-dependent function $\gamma(t)$ introduces additional geometric contributions to the nonmetricity scalar, leading to modified cosmological dynamics compared to the standard coincident gauge formulation. In particular, for the choice $\gamma(t)=H(t)$, the nonmetricity scalar acquires extra terms that persist during the cosmological evolution, allowing for nontrivial deviation from the standard relation $Q \simeq -6H^{2}$.

To analyze the resulting dynamics, we constructed an autonomous system using suitable dimensionless variables and examined two representative $f(Q)$ models: the power-law model and a logarithmic model. In both cases, we identified critical points and studied their stability properties using linear stability theory, along with cosmological parameters such as the deceleration parameter and total EoS parameter.

For the power-law model, the phase space exhibits a cosmological sequence consisting of radiation-dominated, matter-dominated and late-time accelerated phases. In particular, a stable de Sitter attractor emerges, corresponding to a DE-dominated Universe with $\omega_{\text{tot}}=-1$, consistent with observational results. For the logarithmic model, the system also admits a late-time de Sitter attractor and reproduces a cosmological sequence. However, the presence of non-hyperbolic critical points requires a more refined analysis beyond linear stability theory. In particular, we employed CMT to investigate the stability of the critical point $C_2$, confirming its attractor behavior and establishing the robustness of the late-time accelerating solution. This highlights the importance of nonlinear stability techniques in fully characterizing the phase-space structure of the model.

Overall, both models yield a cosmological evolution consistent with observational constraints, smoothly interpolating between radiation, matter and dark-energy-dominated eras, with $\omega_{\text{DE}} \to -1$ late times effectively reproducing $\Lambda$CDM-like behavior. A key outcome of this analysis is that Connection II induces persistent geometric effects that can influence intermediate cosmological dynamics while still allowing a stable late-time acceleration.

\section*{Acknowledgements}
The authors would like to state that this research was conducted without any financial support from public, commercial or nonprofit funding sources.

\bibliographystyle{utphys}
\bibliography{biblio}

\end{document}